\documentclass[onefignum,onetabnum]{siamart171218}

\usepackage{lipsum}
\usepackage{amsfonts}
\usepackage{graphicx}
\usepackage{epstopdf}
\usepackage{algorithmic}
\usepackage{float}
\usepackage{adjustbox}
\usepackage{pdflscape}
\usepackage[T1]{fontenc}
\usepackage[utf8]{inputenc}
\usepackage{booktabs,amsmath,makecell,array,etoolbox}
\makeatletter

\newcolumntype{C}{>{\centering\arraybackslash}X}

\ifpdf
  \DeclareGraphicsExtensions{.eps,.pdf,.png,.jpg}
\else
  \DeclareGraphicsExtensions{.eps}
\fi

\newsiamremark{remark}{Remark}
\newsiamremark{hypothesis}{Hypothesis}
\crefname{hypothesis}{Hypothesis}{Hypotheses}
\newsiamthm{claim}{Claim}

\title{Ill-Conditioning in Dictionary-Based Dynamic-Equation Learning: A Systems Biology Case Study}

\author{
Yuxiang Feng\thanks{
Department of Engineering Sciences and Applied Mathematics,
Northwestern University, Evanston, IL 60208, USA.
}
\and
Niall M. Mangan\footnotemark[1]
\thanks{NSF-Simons National Institute for Theory and Mathematics in Biology, Chicago, IL, USA; Northwestern University, Center for Synthetic Biology, Evanston, IL, USA}
\and
Manu Jayadharan\footnotemark[1] 
\thanks{Author for correspondence. (\email{manu.jayadharan@gmail.com}).
}
}

\usepackage{amsopn}

\makeatletter
\newcommand*{\addFileDependency}[1]{
  \typeout{(#1)}
  \@addtofilelist{#1}
  \IfFileExists{#1}{}{\typeout{No file #1.}}
}
\makeatother

\newcommand*{\myexternaldocument}[1]{%
    \externaldocument{#1}%
    \addFileDependency{#1.tex}%
    \addFileDependency{#1.aux}%
}

\ifpdf
\hypersetup{
  pdftitle={An Example Article},
  pdfauthor={D. Doe, P. T. Frank, and J. E. Smith}
}
\fi

\myexternaldocument{ex_supplement}

\usepackage[
  inner=1in,
  outer=1in,
  top=1in,
  bottom=1in
]{geometry}
\begin{document}

\maketitle

\begin{abstract}
Data-driven discovery of governing equations from time-series data provides a powerful framework for understanding complex biological systems. Library-based approaches that use sparse regression over candidate functions have shown considerable promise, but they face a critical challenge when candidate functions become strongly correlated: numerical ill-conditioning. Poor or restricted sampling, together with particular choices of candidate libraries, can produce strong multicollinearity and numerical instability. In such cases, measurement noise may lead to widely different recovered models, obscuring the true underlying dynamics and hindering accurate system identification. Although sparse regularization promotes parsimonious solutions and can partially mitigate conditioning issues, strong correlations may persist, regularization may bias the recovered models, and the regression problem may remain highly sensitive to small perturbations in the data. We present a systematic analysis of how ill-conditioning affects sparse identification of biological dynamics using benchmark models from systems biology. We show that combinations involving as few as two or three terms can already exhibit strong multicollinearity and extremely large condition numbers. We further show that orthogonal polynomial bases do not consistently resolve ill-conditioning and can perform worse than monomial libraries when the data distribution deviates from the weight function associated with the orthogonal basis. Finally, we demonstrate that when data are sampled from distributions aligned with the appropriate weight functions corresponding to the orthogonal basis, numerical conditioning improves, and orthogonal polynomial bases can yield improved model recovery accuracy across two baseline models.
\newline
\textbf{Relevance to Life Sciences}
Numerical ill-conditioning is especially consequential in the model discovery for biological systems, where nonlinear interactions are often represented using nonlinear functions such as polynomials, and where multiscale dynamics, constrained state trajectories, and limited sampling due to experimental limitations can further amplify multicollinearity. We demonstrate these effects across benchmark models relevant to metabolic networks, regulatory networks, and population dynamics. Our results show that poor conditioning can impair the recovery of biologically meaningful governing equations, while sampling strategies matched to the candidate basis can improve identification accuracy. These results imply that a broader range of dynamic sampling is needed in most biological experiments to produce data sets that are suitable for data-driven model discovery with current methods.
 \newline
\textbf{Mathematical Content}
This paper studies sparse regression-based equation discovery in the presence of multicollinearity and numerical ill-conditioning. We analyze the conditioning of candidate libraries, especially monomial and orthogonal polynomial bases, using condition numbers and model recovery under realistic sampling conditions with publicly available experimental data. We compare how basis choice and sampling distribution affect regression stability, sparsity, and the accuracy of recovered dynamical models.

\end{abstract}

\begin{keywords}
 Data-driven model discovery, Scientific ML, Biological system modeling, Ill-conditioning, Instability, Sparse model selection, AI for system biology, Experimental design.
\end{keywords}


\section{Introduction}
From enzyme kinetics to ecological population dynamics, biological systems are governed by complex networks of interactions that manifest as nonlinear dynamical systems \cite{amaral2004complex, costa2008complex}. Dynamical systems modeling is essential for predicting system behavior, engineering interventions, and understanding which species interactions dominate the physical behavior. Traditionally, such models have been derived by experts and calibrated to data relying on first principles and known biochemical mechanisms \cite{alon2019introduction, klipp2016systems}. However, the increasing availability of time-series data from high-throughput experiments has created opportunities for the data-driven discovery of dynamical models \cite{bar2012studying, rai2023ultra}. Such methods are especially useful when underlying mechanisms are poorly understood or too complex for manual model construction, as is often the case in gene regulatory networks, metabolic pathways, and cellular signaling cascades \cite{bansal2006inference, chandrasekaran2010probabilistic, chou2009recent}. By leveraging data to infer the underlying equations, one can accelerate the discovery of biological mechanisms and capture dynamics that might be missed by human intuition. Data-driven model discovery has advanced over the past decade, with approaches ranging from symbolic regression \cite{schmidt2009distilling} to neural network-based methods \cite{chen2018neural, raissi2019physics}. Equation learning for biological systems has been explored in a range of settings, including differential-equation discovery using stochastic agent-based model simulations \cite{nardini2021learning} and using neural-network-based approaches \cite{lagergren2020biologically,flores2024emb}. Among the available approaches, sparse-regression-based frameworks like SINDy \cite{brunton2016discovering, brunton2016sparse, rudy2017data} and related methods like IDENT \cite{he2022robust,kang2021ident} are directly interpretable and computationally efficient. These approaches formulate model discovery as a sparse optimization problem in which relevant terms are selected from a dictionary or library of nonlinear candidate functions to reconstruct the governing equations. Such approaches have been applied successfully to diverse simulated and, more recently, experimental biological systems \cite{mangan2017inferring,sadria2025discovering, schmitt2024machine, lagergren2020learning}.\par

However, sparse identification methods face fundamental challenges that limit their reliability, especially in biological applications. Noise sensitivity has motivated the development of techniques such as integral and weak formulations \cite{messenger2021weak, reinbold2020using}. Library construction poses additional difficulties, motivating work on automated library learning and physics-informed constraints \cite{champion2020unified, xu2020dlga}. 
An additional challenge is the numerical ill-conditioning arising from multicollinearity among candidate functions 
\cite{belsley2005regression, chan2022mitigating, kaheman2020sindy, kreikemeyer2024discovering}. While diagnostics such as the variance inflation factor (VIF) \cite{o2007caution} are widely used to detect multicollinearity, eliminating highly correlated features is often ineffective in the context of library learning \cite{bao2025information} compared to classical regression settings, since each candidate term typically corresponds to a specific mechanistic interaction in the governing equations and therefore cannot be removed arbitrarily.
Ensemble-based sparse identification approaches have also been proposed to improve robustness in the presence of correlated candidate functions \cite{adrian2024stabilizing}, although these methods address the instability algorithmically rather than resolving the underlying source of feature correlations. Recent studies have highlighted that biological systems, which often evolve on low-dimensional manifolds, naturally induce strong correlations among nonlinear basis functions \cite{delgado2025sindy}. In biological systems, time-scale separation and mass conservation (e.g. enzyme kinetics) often require rational nonlinearities in the candidate library, which increases the non-convexity of the optimization problem involved. A few methods \cite{kaheman2020sindy, mangan2017inferring} address this challenge by reformulating into an implicit equation so that such dynamics can be identified using polynomial libraries. More recently, \cite{jayadharan2025sodas} introduced a sparse-regression framework to jointly discover differential equations, conservation laws, and steady-state approximations, particularly useful for enzyme kinetics, again relying on polynomial functions. 
Polynomial dictionaries are well-suited for systems biology since mass-action kinetics and species interactions naturally take monomial form. The numerical stability of polynomial regression has been extensively studied, where orthogonal polynomial bases are known to provide better conditioning than monomials in polynomial regression \cite{brubeck2021vandermonde, narula1979orthogonal, wishart1953orthogonal} by maintaining orthogonality under specific weight functions, which eliminates cross-correlations. However, this orthogonality requires data to be distributed according to the weight function of each basis \cite{xiu2010numerical}. As we show here, this dependence on the sampling measure has implications for biological systems, where experimental constraints and intrinsic multiscale dynamics rarely produce data conforming to these distributions \cite{reinbold2021robust}. \par

Issues of identifiability have long been recognized in systems biology, most notably through the work on sloppy parameter sensitivities \cite{gutenkunst2007universally}. These insights have motivated a substantial body of research on identifiability studies for systems biology models \cite{chis2011structural, monsalve2022analysis, villaverde2016structural}. In contrast, a comparably systematic study of identifiability and numerical conditioning in biological model discovery using library-based sparse optimization approaches has not yet been undertaken. Here we examine how library ill-conditioning causes instability in sparse identification for biological systems and suggest a principled approach to mitigate these numerical challenges through distribution-aligned sampling. Rather than filtering the candidate library \cite{jayadharan2025sodas} or improving noise robustness \cite{messenger2021weak}, we address ill-conditioning through data distribution alignment. By quantitatively linking library ill-conditioning to failures in equation recovery, we connect modern model-discovery methodologies with classical insights from numerical linear algebra, complementing recent perspectives that emphasize the importance of parsimonious and well-conditioned representations of complex dynamics \cite{brunton2016discovering, kutz2022parsimony, trefethen2019approximation}. Our main contributions include: (i) a comprehensive quantification of the ill-conditioning of the feature library across two baseline models including a Lotka-Volterra (L-V) prey predator model and a chemical reaction network (CRN) and nine benchmark biological models, showing that multicollinearity begins to dominate the feature library as the degree of the polynomial library increases, and that even combinations of two to three features become ill-conditioned in the presence of higher-order monomial terms in the library, (ii) evidence that orthogonal polynomial bases fail to improve the conditioning of the problem when data distributions deviate from theoretical requirements, sometimes performing worse than monomials, and (iii) a demonstration that aligning the sampling distribution with the polynomial basis, implemented through a distribution-aligned sampling strategy, can mitigate numerical conditioning issues and improve recovery of governing equations for the two baseline models. By integrating classical insights about stability of numerical methods with modern data-driven model discovery, this work characterizes fundamental limitations and provides actionable strategies for reliable equation learning in the context of systems biology.

The remainder of this paper is organized as follows. Section 2 presents our main results: Section 2.1 - 2.3 quantify the prevalence of ill-conditioning of the feature matrix across various models in systems biology and analyze the consequences of this poor conditioning for model recovery, Section 2.4 demonstrates the limitations of orthogonal polynomial bases for improving conditioning in practice, and Section 2.5 examines the critical role of data distribution in restoring conditioning and enabling accurate model recovery with orthogonal bases. Section 3 discusses implications of our findings for both theoretical understanding and practical applications of data-driven model discovery for systems biology.


\section{Main results}
\label{sec:main}
Throughout this work, we consider candidate function libraries constructed from polynomial features of the system state variables, following the standard SINDy framework \cite{brunton2016discovering}. The baseline libraries are built using a monomial basis, consisting of all monomials formed from the state variables up to a prescribed maximum degree. For a dynamical system with state vector $\mathbf{x} = (x_1,\dots,x_d)$, a monomial library of degree $p$ includes all terms of the form
\[
x_1^{\alpha_1} x_2^{\alpha_2} \cdots x_d^{\alpha_d},
\]
where the nonnegative integers $\alpha_i$ satisfy $\sum_{i=1}^d \alpha_i \le p$. We refer to this maximum total degree $p$ as the \emph{degree of the library} for the remaining of the paper. \par
We consider dynamical systems of the form
\[
\frac{\mathrm{d}}{\mathrm{d}t}\,\mathbf{x}(t)
= \sum_{l=1}^{\zeta} \xi_l\, f_l\!\big(\mathbf{x}(t)\big),
\]
where the state vector $\mathbf{x}(t)\in \mathbb{R}^n$ denotes the system state at time $t$, and the sum of functions $\sum_{l=1}^{\zeta} \xi_l\, f_l\!\big(\mathbf{x}(t)\big)$ represents the dynamical constraints governing the system's equations of motion. We assume that $\zeta$ is small, reflecting the sparsity of the underlying dynamics. Given a candidate function library $\Theta(\mathbf{x})=\{\theta_1(\mathbf{x}),\theta_2(\mathbf{x})...\theta_p(\mathbf{x})\}$, the functions appearing in the governing equations are assumed to be only a small subset of $\Theta(\mathbf{x})$. Measurements of the state variables are collected into a data matrix $X\in \mathbb{R}^{m\times n}$, where each row is a measurement of the state vector $\mathbf{x}^{\top}(t_i)$ for $i \in \{1,\ldots,m\}$. Evaluating the candidate library on these measurements yields the feature matrix $\Theta(\mathbf{X})\in \mathbb{R}^{m\times p}$. Using time-series data, the active terms in the governing equations are identified by solving the sparse regression problem
\[
\dot{X} = \Theta(X)\,\Xi,
\]
where $\dot{X}\in\mathbb{R}^{m\times n}$ contains the corresponding time derivatives of the state variables, and the coefficient matrix $\Xi = [\,\boldsymbol{\xi}_1\ \boldsymbol{\xi}_2\ \cdots\ \boldsymbol{\xi}_n\,]$ encodes the active nonlinear terms for each state equation.

\subsection{Ill-Conditioning Observed in Baseline Models}
\begin{figure}[h]
    \centering
    \includegraphics[width=0.7\linewidth]{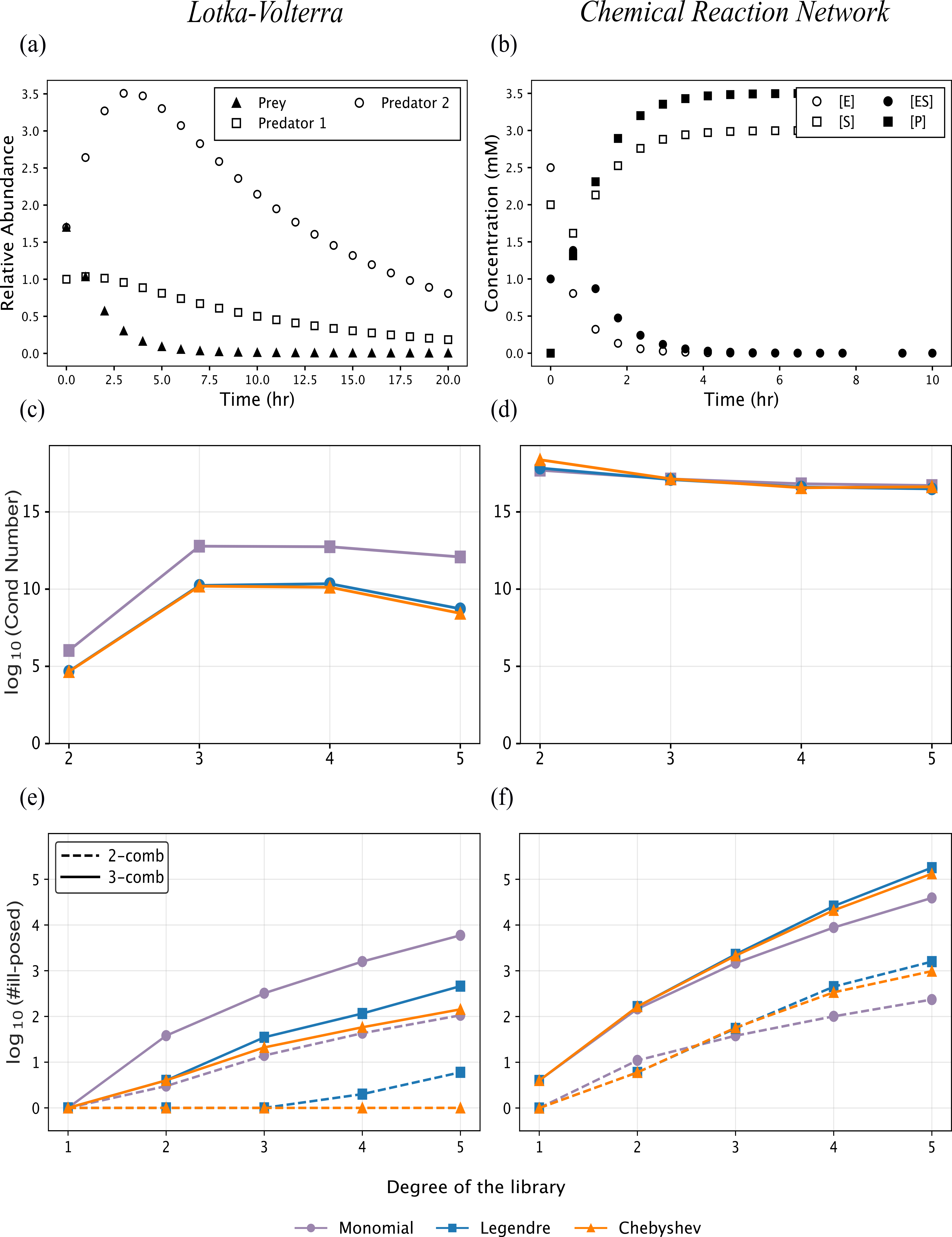}
    \caption{\textbf{Numerical ill-conditioning of candidate function libraries for baseline biological dynamical systems.}
Panels (a)–(b) show representative time-series data generated from numerical simulations of the Lotka–Volterra predator–prey system and a chemical reaction network (CRN), respectively. Panels (c)–(d) report the condition numbers of full polynomial function libraries constructed using monomial, Legendre, and Chebyshev bases as functions of library degree for the two systems. Panels (e)–(f) quantify the prevalence of ill-conditioning by counting the number of ill-conditioned two-term and three-term combinations in the candidate libraries, measured as the base-10 logarithm of the number of combinations exceeding an $\mathrm{R}^2$ threshold. Results are shown separately for the Lotka–Volterra system (left column) and the CRN model (right column).}
    \label{fig:Figure-1}
\end{figure}

A Lotka-Volterra (L-V) system consisting of one prey and two predators species, together with a chemical reaction network (CRN) model comprising four interacting species served as baseline models in this study. The governing equations of these two baseline models can be found in \cref{Table:Table-1}. Synthetic time-series data were generated through numerical simulation following realistic sampling rates found in \cite{kennedy2022linking, odum1971fundamentals} (\Cref{fig:Figure-1}(a) and \Cref{fig:Figure-1}(b)). \par

The condition number of the matrix representing the candidate library, where the columns form the features, provides a useful measure of the conditioning of the regression problem encountered during model discovery \cite{kim2019multicollinearity, mangan2019model}. \Cref{fig:Figure-1}(c) and \Cref{fig:Figure-1}(d) compare the condition numbers of feature libraries constructed using a monomial basis with varying highest-degree. The figure demonstrates that the condition number remains extremely high across different degree of the feature candidate library. However, the condition number aggregates multicollinear relationships among features in the library and generally increases with the overall number of features. To inspect the number of features required to form a linearly dependent set, we analyzed the number of ill-posed two- and three-term feature combinations corresponding to pairwise and three-way dependencies, respectively, in baseline models constructed using the monomial basis: these results are reported in \Cref{fig:Figure-1}(e)--\Cref{fig:Figure-1}(f), respectively. There are strong multicollinear relationships between two and three terms in the library, and the number of relationships increases with the polynomial degree, with multicollinearity becoming substantially more prevalent in higher-degree libraries. This behavior reflects the combinatorial growth of candidate functions and the rapidly expanding space of possible term interactions \cite{pan2016bad}. 
Consequently, the monomial basis becomes less stable as the degree of the candidate library increases, indicating that numerical ill-conditioning is an intrinsic property of high-degree polynomial libraries rather than a consequence of a few isolated feature interactions.

\subsection{Ill-Conditioning Contributes to Model Misidentification} 

To understand the driving factors behind the misidentification of the model by sparse-regression methods, we analyzed the pattern of model misspecification across the baseline models and different library degrees. We compared recovered and true equations term-by-term and observed a consistent mechanism: terms that appear incorrectly in the identified model (false positives) tend to be strongly correlated with the true terms that are missing (false negatives). We quantified this by forming, for each equation, a matrix from the candidate functions corresponding to the union of incorrectly selected terms and missing true terms, and then evaluating its condition number (\Cref{Table:Table-1}). This provides an after-the-fact diagnostic of whether the specific substitutions made by the regression are occurring in a numerically ill-posed subspace.\par

\Cref{Table:Table-1} presents a term-wise comparison between the ground-truth governing equations and the expressions recovered by SINDy \cite{brunton2016discovering} for the L-V and CRN systems. All model discovery experiments were conducted using the PySINDy package \cite{de2020pysindy}, a widely used implementation of different variants of SINDy. Notably, the numerical conditioning issues and recovery failures observed are not specific to SINDy and are expected to arise in other library-based methods as well. In experiments on L-V and CRN, the recovered models exhibit the same failure pattern: when a true term is missing from the discovered model, new terms are introduced whose corresponding features are highly correlated with the missing term. The conditioning analysis explains this behavior. When the combined set of missing and erroneously included terms is ill-conditioned, the candidate functions become strongly correlated, and the regression cannot uniquely attribute contributions to the correct features. In this regime, a missing true term can be replaced by a linear combination of correlated alternatives, leading to systematic structural errors. For the Lotka-Volterra system, the condition number of the submatrix formed by the incorrectly identified terms reaches $\mathcal{O}(10^{5})$, consistent with the full library conditioning $\mathcal{O}(10^{6})$ and coinciding with an inability to distinguish among correlated nonlinear terms. The CRN system shows the same phenomenon with the sub-library and full library having conditioning $\mathcal{O}(10^{17})$. It appears that large condition numbers of the overall library arise, at least in part, from the subset of highly correlated terms that are frequently misidentified. 

\begin{table}[h]
  \centering  
  \renewcommand{\arraystretch}{1.05}
  \setlength{\tabcolsep}{1.5pt} 
  \small
  \caption{\textbf{Conditioning of error-associated subspaces in sparse identification of baseline dynamical systems.}
For each equation of the Lotka–Volterra (L-V) and chemical reaction network (CRN) baseline models, the table lists the true governing equation and the equation recovered by SINDy using a monomial polynomial library (degree 3 for L-V and degree 2 for CRN), together with the sets of false negatives or missing and false positives or wrong terms. The reported condition number corresponds to the submatrix formed by the candidate functions associated with these misidentified terms. }
  \label{Table:Table-1}

  \begin{tabular}{@{} 
    c 
    r@{}l   
    r@{}l   
    c c c 
    @{}
}
\toprule
System & 
\multicolumn{2}{c}{\makecell{True\\Equation}} & 
\multicolumn{2}{c}{\makecell{Discovered\\Equation}} & 
\makecell{Missing\\Terms} & 
\makecell{Wrong\\Terms} & 
\makecell{Con([Missing,\\Wrong])} \\
\midrule

LV (Eq1) &
$\dot x_1$ & 
$\begin{aligned}[t]
  {}= &\, 0.2 x_1 - 0.118 x_1^2 \\
      &\, - 0.1 x_1x_2 - 0.2x_1x_3
\end{aligned}$ &
$\dot x_1$ & 
${}= 0.119 x_1^3 - 0.273 x_1^2 x_3$ &
\makecell{$x_1,\, x_1x_2,$ \\ $ x_1x_3,\, x_1^2$} &
$x_1^2x_3,\, x_1^3$ &
$8.13e5$ \\
\addlinespace[3pt]

LV (Eq2) &
$\dot x_2$ & 
${}= -0.1x_2 + 0.1x_1x_2$ &
$\dot x_2$ & 
$\begin{aligned}[t]
  {}= &\, -0.023 x_3 + 0.031 x_1x_3 \\
      &\, + 0.015 x_1^2 x_3 - 0.02 x_1x_2x_3
\end{aligned}$ &
$x_2,\, x_1x_2$ &
\makecell{$x_3,\, x_1x_3,$ \\ $x_1x_2x_3,\, x_1^2x_3$} &
$3.04e3$ \\
\addlinespace[3pt]

LV (Eq3) &
$\dot x_3$ & 
${}= -0.1x_3 + 0.4x_1x_3$ &
$\dot x_3$ & 
$\begin{aligned}[t]
  {}= &\, -0.098x_3 + 6.136 x_1^2 \\
      &\, - 2.02 x_1^3 - 1.441 x_1^2x_3\\
      &\, + 0.109 x_1x_3^2
\end{aligned}$ &
$x_1x_3$ &
\makecell{$x_1^2,\, x_1x_3^2,$ \\ $x_1^2x_3,\, x_1^3$} &
$1.23e5$ \\
\addlinespace[6pt]

CRN (Eq1) &
$\dot x_1$ & 
${}= 1.10x_3 - 2.12x_1x_3$ &
$\dot x_1$ & 
$\begin{aligned}
    {}= &\,  -1.773 x_1 + 0.664 x_1^2 \\
        &\,  + 0.518 x_3^2
\end{aligned}$ &
$x_3,\, x_1x_2$ &
$x_1,\, x_1^2,\, x_3^2$ &
$1.36e17$ \\
\addlinespace[3pt]

CRN (Eq2) &
$\dot x_2$ & 
${}= 2.61x_3 - 2.12x_1x_2$ &
$\dot x_2$ & 
$\begin{aligned}[t]
  {}= &\, 0.485x_1^2 - 1.538 x_1x_3 \\
      &\, - 0.636 x_2x_3 + 0.672x_3^2\\
      &\, + 0.907 x_3x_4
\end{aligned}$ &
$x_3,\, x_1x_2$ &
\makecell{$x_1x_3,\, x_2x_3,$ \\ $x_3x_4,\, x_1^2,\, x_3^2$} &
$2.00e18$ \\
\addlinespace[3pt]

CRN (Eq3) &
$\dot x_3$ & 
${}= 2.12x_1x_2 - 2.61x_3$ &
$\dot x_3$ & 
$\begin{aligned}[t]
  {}= &\, -0.485x_1^2 + 1.538 x_1x_3 \\
      &\, + 0.636 x_2x_3 - 0.672x_3^2 \\
      &\, - 0.907 x_3x_4
\end{aligned}$ &
$x_3,\, x_1x_2$ &
\makecell{$x_1x_3,\, x_2x_3,$ \\ $x_3x_4,\, x_1^2,\, x_3^2$} &
$2.00e18$ \\
\addlinespace[3pt]

CRN (Eq4) &
$\dot x_4$ & 
${}= 1.51x_3$ &
$\dot x_4$ & 
$\begin{aligned}[t]
  {}= &\, 0.106 x_3 + 0.367 x_1 x_2 \\
      &\, + 0.033 x_1 x_3 + 0.309 x_1 x_4 \\
      &\, + 0.155 x_2 x_3 + 0.163 x_3^2 \\
      &\, + 0.175 x_3 x_4
\end{aligned}$ &
 & 
\makecell{$x_1 x_2,\, x_1 x_3,\, x_1 x_4,$ \\ $x_2 x_3,\, x_3 x_4,\, x_3^2$} &
$2.566e4$ \\
\addlinespace

\bottomrule
\end{tabular} 
\end{table}

\subsection{Ill-Conditioning is Prevalent in a Wide Range of Systems Biology Models} 
Having demonstrated the presence of severe ill-conditioning in baseline models and its possible detrimental impact on sparse model identification, we now show that this issue is not isolated but is widespread across systems biology by examining a collection of benchmark models in this section. \par
Our collection of systems biology models was primarily adapted from \cite{hass2019benchmark}. These models cover a wide spectrum of biological systems, including metabolic networks, regulatory networks, and population dynamics. Each model is formulated as a system of ordinary differential equations (ODEs), with the right-hand side expressed in polynomial form. The number of state variables ranged from four to fifteen which can be used as a measure of model complexity. The dataset in \cite{hass2019benchmark} included both experimental measurements and interpolated data, with the interpolated points generated from the experimental measurements to compensate for sparsely sampled time intervals. We conducted numerical experiments on the interpolated sets, as they preserve the experimentally observed dynamic range and dynamic features of the system while mitigating issues associated with low temporal resolution that can compromise derivative estimation in data-driven model discovery.\par

\begin{figure}[h]
    \centering
    \includegraphics[width=1.0\linewidth]{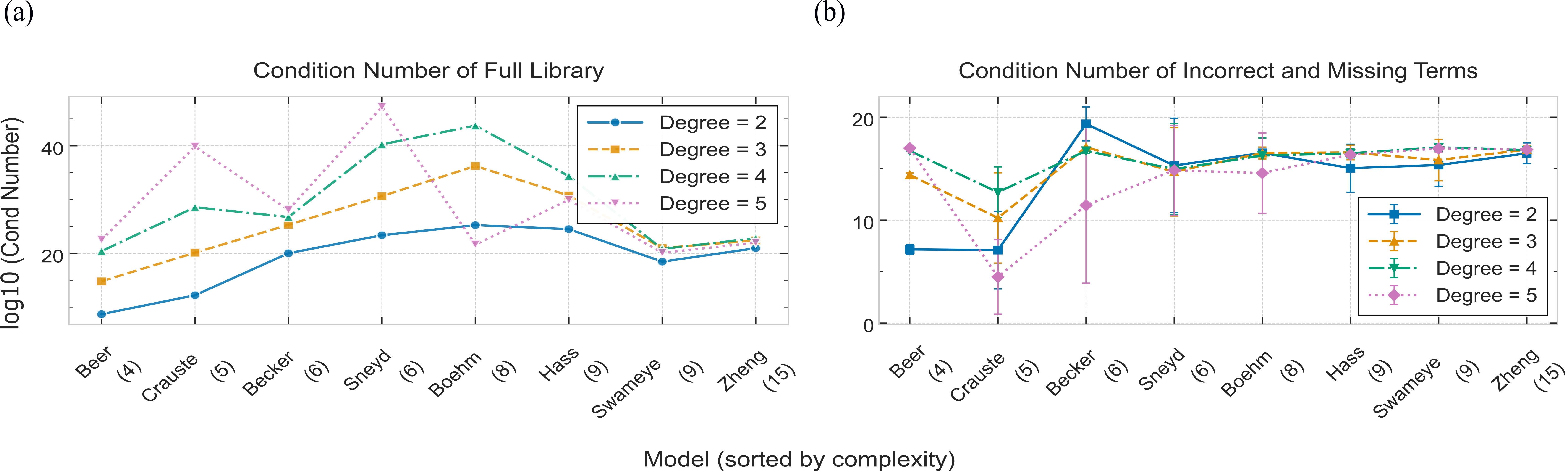}
    \caption{\textbf{Conditioning of candidate function libraries across benchmark biological models as a function of model complexity.}
 Panel (a) shows the dependence of the full library condition number on model complexity for each model. Panel (b) focuses on the conditioning of sub-matrices formed by features corresponding to false negatives or missing and false positives or wrong terms in the mis-identified model after sparse regression, with each point representing the mean across equations within a model and error bars indicating one standard deviation.}
    \label{fig:Fig-4}
\end{figure}

The relationship between model complexity and library conditioning is illustrated in \Cref{fig:Fig-4}. Model complexity is quantified by the number of states in each system. While models with higher state dimensions exhibit consistently large condition numbers for a fixed polynomial degree, there is no clear monotonic behavior in the condition number as the number of state variables increases. Examination of the condition numbers of the full function library across benchmark biological models can be found in supplementary fig SM1(a) (exact values can be found in supplementary table SM1)
showed that generally, the condition number increased as the library polynomial degree is raised from 2 to 5, reflecting the expected growth in multicollinearity induced by higher-order polynomial terms. 
This suggests that the ill-conditioning of nonlinear function libraries is not determined solely by polynomial degree or number of state variables, but is also strongly influenced by the geometric and dynamical complexity of the underlying system. In particular, systems with multiscale reaction dynamics or nested signaling architectures tend to produce state trajectories confined to low-dimensional manifolds \cite{lee2010multi, gorban2018model}. Such concentration enhances collinearity among nonlinear candidate functions, thereby amplifying library ill-conditioning and reducing the robustness of sparse model recovery. 

\Cref{fig:Fig-4}(b) presents a complementary perspective by analyzing the condition numbers of matrices formed by incorrectly identified terms (both wrong and missing terms) during the recovery process (see also supplementary fig SM1(b); exact values can be found in supplementary table SM2). Unlike the full-library analysis, no consistent dependence on polynomial degree is observed. When viewed as a function of model complexity (\Cref{fig:Fig-4}(b)), the condition numbers associated with the error-term subspaces exhibit a modest upward trend, albeit with notable deviations. In particular, a pronounced peak appears at an intermediate complexity level, corresponding to the Becker model, indicating a regime in which the sparse regression problem becomes especially ill-posed. The mean condition numbers of these error-associated matrices remain uniformly high and largely insensitive to increases in library degree.
\par

\subsection{Why Orthogonal Bases Do Not Automatically Improve Conditioning in Practice}
Orthogonal polynomial bases are known to transform an otherwise ill-conditioned system into a well-conditioned one \cite{gautschi1983condition}. By ensuring that the basis functions used to construct the features are linearly independent, orthogonality minimizes cross-correlations among library columns even as the number of features increases. This substantially mitigates the ill-conditioning, particularly the numerical instability in polynomial regression \cite{brubeck2021vandermonde, narula1979orthogonal}. As a result, classical statistical learning literature predicts a dramatic reduction in condition numbers and a numerically well-posed regression problem \cite{gautschi1983condition, higham2002accuracy}, and orthogonal polynomials are often proposed in the sparse model identification literature as a potential way to mitigate correlation \cite{roman2025approximating}\par

Despite these theoretical advantages, orthogonal bases do not substantially alleviate ill-conditioning in practice for library-based regression methods when experimental limitations restrict sampling, as is often the case in biological systems such as those considered in this study.  For both baseline models, nonlinear function libraries constructed from orthogonal polynomials still exhibit markedly large condition numbers (\Cref{fig:Figure-1}(c) and \Cref{fig:Figure-1}(d)), directly contrasting with the classical conditioning advantages predicted by the orthogonality of these specialized bases. This conclusion is supported by the analysis of the number of ill-posed combinations and the condition number across the orthogonal basis libraries, including the Legendre, Chebyshev bases. These bases are classical and widely used orthogonal polynomial families  \cite{shen2011spectral}. Similar behavior is observed in the analysis of the Laguerre basis presented in supplementary section SM4. For the simpler Lotka-Volterra system, orthogonal bases provide only a modest reduction in the number of ill-posed combinations relative to the monomial basis, and this advantage deteriorates rapidly as the polynomial degree increases, with ill-posed combinations growing across all bases. The benefits of orthogonality are shown to be negligible at higher degrees. A notable exception is the degree-2 Chebyshev basis shown in \Cref{fig:Figure-1}(e), for which no ill-posed two-term combinations are detected, showing some pair-wise de-correlation was achieved at the lowest degree, although this advantage is not guaranteed to persist at higher degrees. This degradation is further amplified when extending the analysis from pairwise dependencies to three-way interactions, indicating that multicollinearity starts building up within the feature matrix even at the level of linear combination of three features even though de-correlation is expected from orthogonality. The effect is even more pronounced for the more complex CRN. In this case, the overall level of ill-posedness is substantially higher across all bases, and orthogonal polynomial libraries do not consistently outperform monomials. In fact, for high-degree libraries and for three-term combinations, orthogonal bases can exhibit even stronger collinearity than the monomial basis. 

Together, these results show that the orthogonal library features exhibit high multicollinearity, even in small term combinations. This suggests that features constructed from orthogonal polynomial bases are not truly orthogonal in practice. As a result, the theoretical numerical advantages promised by these bases are not fully realized in the regression setting. These limitations arise because the benefits of orthogonality depend critically on the underlying data distribution: orthogonality is preserved only when the data are sampled according to the weight function that define the basis. In the absence of distribution-aware sampling or experimental design that enforces these measures, orthogonal polynomial bases do not automatically yield improved conditioning in model discovery, particularly for complex biological systems.

\subsection{Conditioning and Model Recovery Improves with Sampling from Appropriate Distributions}

The lack of improvement in the conditioning of the problem (\Cref{fig:Figure-1}(c)-(f)) after switching to orthogonal bases can be largely attributed to the dependence of orthogonality on the underlying data distribution. Theoretically, orthogonal polynomial bases preserve orthogonality only when the data are sampled according to the specific weight functions or domains under which the polynomials are defined \cite{szeg1939orthogonal}. Deviations from these theoretical distributions disrupt orthogonality among basis functions, reintroducing correlations and numerical instabilities \cite{gautschi1982generating}. Therefore, maintaining orthogonality in practice requires that sampled data conform to distributions consistent with the corresponding orthogonal polynomials. To explicitly test the hypothesis that mismatches between the empirical data distribution and the theoretical distribution implied by the orthogonality weight function drive the observed loss of numerical conditioning and model recovery, we analyzed how deviations from the ideal data sampling distribution, arising from experimental data-collection limitations or intrinsic properties of the dynamical system, affect the conditioning and stability of orthogonal basis libraries.\par

\begin{figure}
    \centering
    \includegraphics[width=0.8\linewidth]{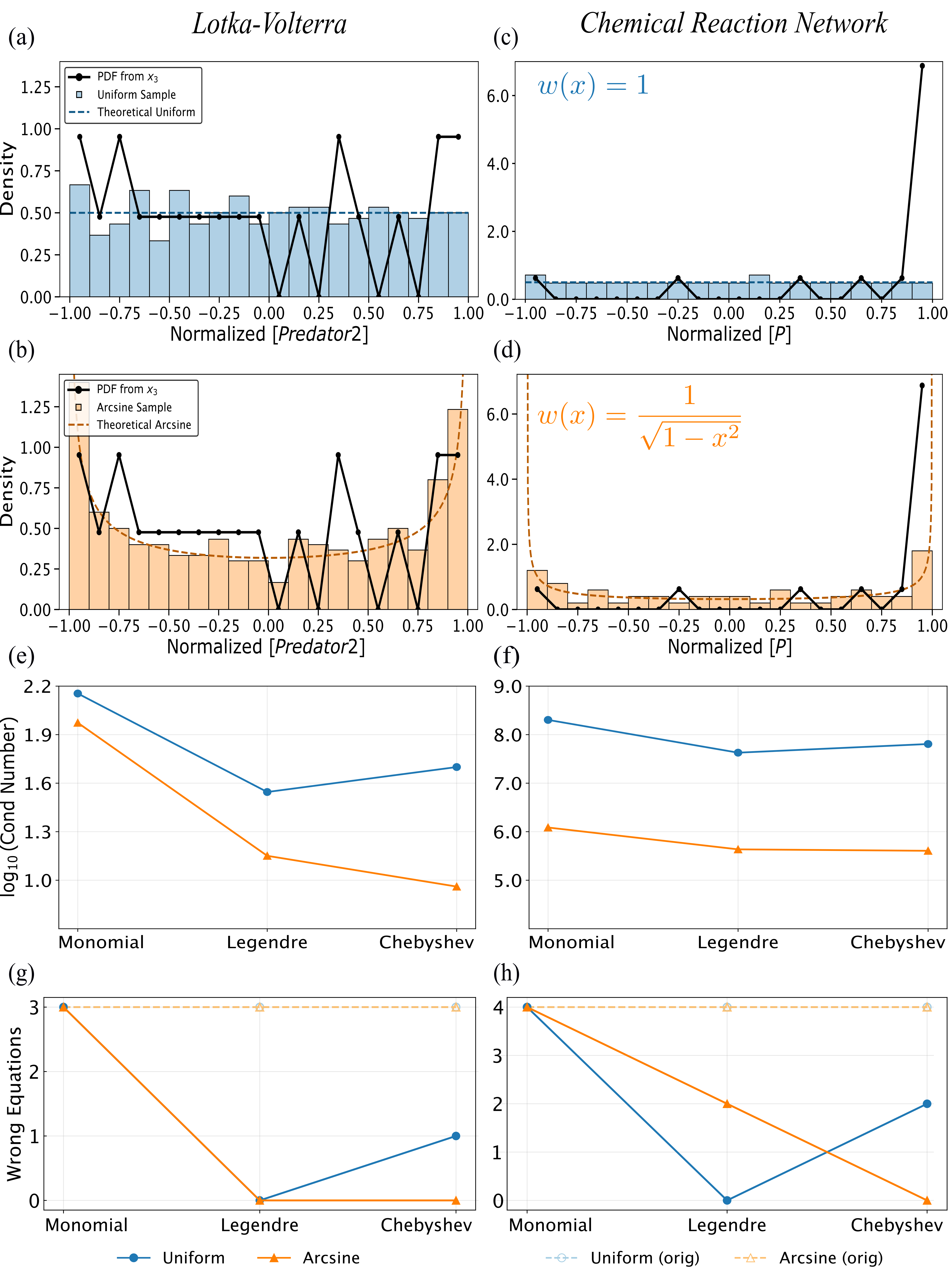}
    \caption{\textbf{Distribution–basis alignment restores orthogonality, improves conditioning, and enables accurate sparse model recovery.}
Panels (a)–(d) compare the empirical distributions induced by system dynamics with the theoretical distributions associated with the weight functions required to preserve orthogonality for Legendre and Chebyshev bases. Solid curves represent empirical probability density functions constructed from simulated state trajectories generated by numerical simulation of the L-V system (left column) and a CRN (right column), reflecting the data distributions induced by the underlying dynamics and serving as proxies for experimentally observed data. Histograms depict data resampled from these trajectories according to the corresponding idealized distributions implied by the orthogonality weight functions (uniform for Legendre and arcsine for Chebyshev). Panels (e)–(f) report the condition numbers of full candidate function libraries constructed from monomial and orthogonal polynomial bases when evaluated on original data versus distribution-aligned samples. Panels (g)–(h) show the resulting model identification errors quantified as the number of incorrectly recovered equations. Candidate function libraries in panels (e)–(h) are constructed using polynomial bases of degree 5.}
    \label{fig:Figure-3}
\end{figure}


\Cref{fig:Figure-3} compares and highlights the discrepancy between dynamical data-induced distributions and the theoretical sampling measures required to preserve orthogonality. In practice, it is often difficult to obtain data that follow the theoretical distributions required to maintain orthogonality either due to experimental or physical constraints. Experimental data are typically affected by measurement noise, uneven sampling, or limited observation ranges, while the dynamical behavior of the system may not naturally conform to the weight functions associated with orthogonal polynomials. Here, we consider a data sampling strategy designed to enforce specific target distributions while preserving the intrinsic dynamics of the underlying ODE system as much as possible. In order to achieve this, initial conditions are sampled from a hypercube, and short trajectories from each initial condition are simulated subsequently. Uniform coverage is achieved using Sobol' quasi-random sequences \cite{burhenne2011sampling}, followed by subsampling strategies to improve spatial uniformity. More general target distributions are obtained via appropriate transformations. The detailed sampling procedure is provided in supplementary section SM5. In principle, this sampling procedure can be applied to systems in benchmark models, but extending it to large-scale systems from real biological experiments is often impractical. \par


\Cref{fig:Figure-3}(e) and \Cref{fig:Figure-3}(f) show that when time-series data are sampled from appropriate distributions in which they are orthogonal, the resulting feature libraries exhibit substantially improved numerical conditioning for both the Lotka–Volterra and CRN models. This improvement in conditioning coincides with more reliable model recovery using SINDy, with perfect recovery for the two baseline models. In principle, orthogonality and numerical conditioning can be further improved by increasing data coverage through additional initial conditions and longer trajectories. However, our results indicate that exact orthogonality is not strictly necessary for successful model reconstruction: even partial or approximate orthogonality is sufficient to enable accurate recovery in the examples considered here. The corresponding sampling patterns in the state space are illustrated in supplementary section SM7.2. This observation is particularly relevant for experimental design in systems biology, where data acquisition is often constrained, and it is rarely feasible to collect large datasets that precisely conform to a prescribed sampling distribution. Instead, modest improvements in distributional coverage achieved through carefully chosen experimental conditions may already yield substantial gains in numerical stability and model identifiability. The Legendre basis show comparably good conditioning under both uniform and arcsine distributions. This behavior can be attributed to the similarity between empirical histograms of the arcsine and uniform distributions, particularly when the number of sampled data points is limited, leading to approximate preservation of orthogonality. Moreover, the CRN model exhibits slightly higher condition numbers than the Lotka–Volterra system across all bases. This difference likely reflects a combinations of factors, including stronger multicollinearity arising from increased system size, as well as the more restricted exploration of state space induced by the dynamics of the CRN, which limit the diversity of observed trajectories. Such constrained dynamics can exacerbate conditioning issues but also make more drastic improvements from distribution-aligned sampling. \par

We assessed SINDy’s model recovery performance using orthogonal basis libraries constructed from data sampled according to the basis-specific theoretical weighting distributions. The orthogonality of the Legendre and Chebyshev bases is defined on the interval [-1,1], therefore, each state variable is linearly rescaled to this interval prior to constructing the corresponding function libraries. \Cref{fig:Figure-3}(g) and \Cref{fig:Figure-3}(h) demonstrate that aligning the sampling distribution with the orthogonality weight of the basis substantially enhances SINDy’s ability to identify the correct model structure, leading to perfect model recovery in the baseline cases. This improvement occurs in regimes where the corresponding libraries constructed from the original data distributions exhibited large condition numbers (exceeding $\mathcal{O}(10^{8})$ for the Lotka-Volterra system and $\mathcal{O}(10^{16})$ for the CRN model), indicating severe ill-conditioning prior to the distribution alignment. 
We emphasize that numerical conditioning serves as a trend-level predictor of recovery performance rather than as an absolute criterion, which means there is no universal threshold in the condition number that guaranties successful model identification \cite{mangan2019model}. The comparable performance (zero identification errors) of the Legendre basis under both uniform and arcsine sampling distributions reinforces the earlier observation that, when the number of sampled data points is limited, the empirical histograms of these two distributions exhibit strong similarity. This resemblance allows the Legendre basis to approximately preserve its orthogonality, thereby maintaining numerical stability and achieving high model identification accuracy. Overall, the results indicate that using orthogonal features together with an appropriate sampling distribution can improve the conditioning of the feature library and, at the same time, enhance the performance of sparse-regression-based model discovery, such as SINDy. We also note that increased model complexity, as exemplified by the CRN, can amplify the performance degradation caused by mismatched sampling distributions.

\section{Discussion}
By examining benchmark models from the systems biology literature, along with two popular baseline models, this work demonstrated a fundamental challenge in the sparse identification of biological dynamical systems: the pervasive ill-conditioning arising from multicollinearity in nonlinear polynomial function libraries substantially impairs model recovery. We argued that orthogonal polynomial bases fail to consistently mitigate ill-conditioning despite their theoretical advantages in de-correlating the candidate functions. Indeed, the orthogonal libraries showed even greater collinearity than the monomials at high degrees or for multi-term interactions in certain cases. Consequently, library-based model discovery methods like SINDy failed to recover the correct model structure under these ill-conditioned settings, omitting important interaction terms and instead selecting spurious high-order terms that correlate with the instabilities. Ill-conditioning thus causes identification errors, a major challenge for biological model discovery. To exploit the conditioning benefits offered by orthogonal polynomial bases relative to standard monomials, we designed a distribution-aligned sampling scheme specific to each basis and analyzed how this alignment influences the accuracy and stability of model recovery. For both the L-V system (one prey and two predators) and a representative CRN model, we verified that achieving the appropriate data distribution for each orthogonal basis is essential for ensuring orthogonality and subsequently enabling accurate recovery of the correct model structure. These results demonstrate a theoretical fix with baseline models to this challenge among biological systems. In practical biological applications, effective model discovery requires that careful experimental design be coupled with an informed choice of function bases, as neither alone is sufficient.\par


Data-driven model discovery methods infer both the connectivity and dynamics of a physical system from the choice of library and information contained in the data. Consequently, the quality, diversity, and geometric structure of the data are critical, as the recovered governing equations are effectively determined by the information encoded in the observed trajectories. Biological systems, however, often exhibit complex, multiscale dynamical characteristics that naturally violate the distributional assumptions required for numerical stability in orthogonal bases. This provides a mechanistic explanation for why data-driven discovery of biological models remains challenging even with intentionally chosen orthogonal bases. In practice, this means a data-driven analysis might falsely dismiss a real regulatory interaction or suggest a fictitious nonlinear effect purely due to numerical instability rather than actual biology. For researchers seeking to infer mechanisms from data, our results highlight the importance of careful experiment design and sufficient data diversity. Ensuring that measurements probe a broad range of system behavior (for example, through varied initial conditions) helps reduce correlations among candidate features and improves the robustness of sparse model identification. In our numerical tests, improved conditioning translated directly into more accurate recovery of the true model, demonstrating that aligning experimental strategies with mathematical requirements (such as approximating the weight distributions corresponding to each orthogonal basis) can substantially enhance model identifiability. Incorporating these numerical considerations into systems biology workflows will help ensure that models inferred from data remain both computationally stable and biologically meaningful.\par

In this study, we primarily analyzed simulated noise-free data for baseline models and interpolated experimental data for benchmark models to approximate experimental sampling conditions. However, experimental measurements, especially in biological systems, are often noisier, more irregularly sampled, and more restricted in dynamical range, which can make model identification substantially more challenging. Furthermore, both the benchmark and baseline systems examined here were assumed to possess dynamics that are well approximated by polynomial representations. If the concerned biological processes involve nonlinearities that are poorly captured by polynomial bases, the resulting library may be misspecified, motivating the use of more general function libraries or alternative sparse-regression based methods. In addition, we did not systematically compare the performance of different sparse regression solvers under severe ill-posed scenarios. While these considerations lie beyond the scope of the present study, they also highlight promising directions for future work. Designing candidate function libraries that are better suited to a given experimental sampling of data is a subject of our current research, and approaches that can guide experimental design to mitigate conditioning and identifiability challenges can provide especially valuable tools for biological model discovery. Progress in these directions could help make data-driven discovery methods more robust, practical, and informative.



\section*{Acknowledgments}
In preparing this manuscript, the authors used OpenAI’s GPT-5 and Grammarly exclusively for language editing, including spelling, grammar, and general stylistic polishing. No AI tools were used for scientific interpretation, analysis, or conclusions, and the authors take full responsibility for the manuscript’s content.

\section*{Declarations}

\paragraph{\textbf{Funding}}
 N.\ M.\ M.\ and M.\ J.\ were supported primarily by the U.S.\ Department of Energy, Office of Science, Office of Advanced Scientific Computing Research, under Award Number DE-SC0024253. Any opinions, findings, conclusions, or recommendations expressed in this material are those of the author(s) and do not necessarily reflect the views of the U.S.\ Department of Energy. Y. F. N.M.M. and M.\ J.\ were additionally supported by the National Institute for Mathematics and Theory in Biology (Simons Foundations award MPS-NITMB-00005320 and National Science Foundation award DMS-2235451).

\paragraph{\textbf{Conflict of interest}}
The authors declare that they have no conflict of interest.

\paragraph{\textbf{Data and Code availability}}
The data and code used to support the findings of this study is publicly available in the project GitHub repository: \href{https://github.com/mjayadharan/IllposedLearning}{\texttt{https://github.com/mjayadharan/IllposedLearning}}.
\paragraph{\textbf{Author contributions}}
M.J conceived the study. M.J and N.M.M designed the research. Y.F performed the analysis and generated the results. Y.F, M.J and N.M.M interpreted the results. N.M.M procured funding for the research. Y.F drafted the initial manuscript, and all authors reviewed, edited, and approved the final manuscript.

\bibliographystyle{siamplain}
\bibliography{references}

\end{document}